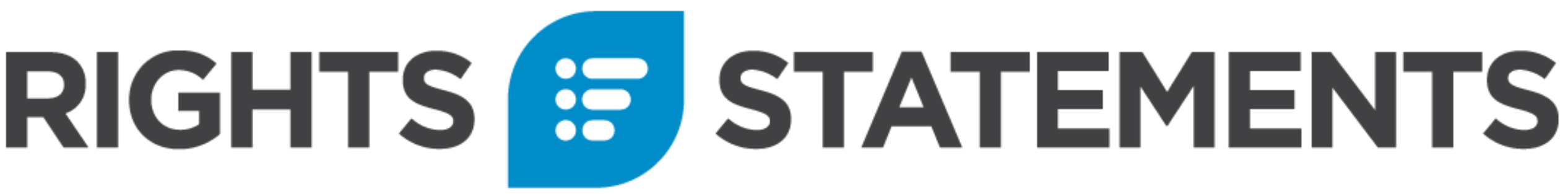

# RightsStatements.org White Paper:

# Requirements for the Technical Infrastructure for Standardized International Rights Statements

RightsStatements.org, October 2015
(updated April 2020)

# RightsStatements.org Technical Working Group

This document is edited by the Technical Working Group of the RightsStatements.org consortium, co-chaired by **Mark A. Matienzo**, Assistant Director for Digital Strategy and Access for Stanford University Libraries, and **Antoine Isaac**, R&D Manager for Europeana, with members:

**Sascha Adler**, Lead Application Developer, Canadian Research Knowledge Network
**Plaban Kumar Bhowmik**, Assistant Professor, Indian Institute of Technology Kharagpur
**Valentine Charles**, Data R&D Coordinator, Europeana
**Esmé Cowles,** Digital Infrastructure Developer, Princeton University Library
**Karen Estlund**, Associate Dean for Technology and Digital Strategies, Penn State University Libraries
**Tom Johnson**, Senior Digital Library Systems Engineer, University of California Santa Barbara Libraries
**Patrick Peiffer**, Digital Librarian, Bibliothèque Nationale de Luxembourg
**Mark Raadgever**, Trove data team, National Library of Australia
**Richard J. Urban**, Digital Asset Manager and Strategist, Corning Museum of Glass
**Maarten Zeinstra**, Advisor, Copyright and Technology, Kennisland

# Introduction

This document is part of the deliverables created by the RightsStatements.org consortium. It provides the technical requirements for implementation of the Standardized International Rights Statements. These requirements are based on the principles and specifications found in the normative *Recommendations for Standardized International Rights Statements*.[1]

This document replaces and supersedes the previously released version of this white paper[2] and the draft white paper, *Recommendations for the Technical Infrastructure for Standardized Rights Statements*, prepared by this working group.[3]

The *Requirements for the Technical Infrastructure for Standardized International Rights Statements* describes the expected behaviours for a service that enables the delivery of human and machine-readable representations of the Rights Statements (note that since the first version of this document, the service has been implemented and deployed at RightsStatements.org). It documents the fundamental decisions that informed the development of a data model grounded in Linked Data approaches. This document also provides proposed implementation guidelines and a non-normative set of examples for incorporating Rights Statements into provider metadata.

The keywords **MUST**, **MUST NOT**, **REQUIRED**, **SHOULD**, **SHOULD NOT**, **RECOMMENDED**, **MAY**, and **OPTIONAL** used in this document are to be interpreted as described in RFC 2119[4].

In this document, Rights Statements (capitalized) refers to the rights statements published by RightsStatements.org.

[1] http://rightsstatements.org/files/160208recommendations_for_standardized_international_rights_statements_v1.1.pdf

[2] http://rightsstatements.org/files/151002requirements_for_the_technical_infrastructure_for_standardized_international_rights_statements.pdf

[3] http://rightsstatements.org/files/150701_recommendations_for_technical_infastructure_for_standardized_international_rights_statements.pdf

[4] https://www.ietf.org/rfc/rfc2119.txt

# URI Design

The infrastructure for RightsStatements.org will host persistent, dereferenceable URIs that enable the delivery of human and machine-readable representations of the Rights Statements.

The URIs that serve as the primary identifiers of the Rights Statements themselves break down into the following components:[5]

| Component | Example Value | Notes |
|---|---|---|
| URI base/domain name | `http://rightsstatements.org/` | **REQUIRED** |
| Resource type | `vocab` | **REQUIRED** |
| Identifier for the statement | `NoC-CR` | **REQUIRED** |
| Version of the statement | `1.0` | **REQUIRED** |

All Rights Statement URIs **MUST** have a trailing slash.

Following these guidelines, the URI for the “No Copyright - Contractual Restrictions (NoC-CR)” Rights Statement is:

`http://rightsstatements.org/vocab/NoC-CR/1.0/`

A machine-readable description of each statement **MUST** be made available via the URI above, following the requirements and recipes presented in the sections “Data Modelling,” “Publication and Implementation,” and the subsections that immediately follow this one.

## URI patterns

Following the recommendations set out in the *Best Practice Recipes for Publishing Linked Data Vocabularies*, we specify the following URI patterns:

[5] Note that the presence of these components in the URI (for example, "NoC-CR" or "1.0") **SHOULD NOT** be considered as an expression of machine-readable metadata. Corresponding facts still need to be asserted at the data level (e.g. there **MUST** be an RDF triple that asserts the version of the Rights Statement to be 1.0). We consider URIs to be opaque for machine interpretation.

| Resource type | Pattern and example URIs without base URI<br>**$variable** denotes variables; [] denotes **OPTIONAL** components |
|---|---|
| Rights Statement / vocabulary concept (vocab) | /vocab/**$id**/**$version**/<br>/vocab/InC/1.0/<br>/vocab/NoC-CR/1.0/ |
| Machine-readable RDF representation (data) | /data/**$id**/**$version**[.***$extension***]<br>/data/InC/1.0/<br>/data/InC/1.0.ttl<br>/data/NoC-CR/1.0/<br>/data/NoC-CR/1.0.json |
| Human-readable representation (i.e. HTML versions) (page) | /page/**$id**/**$version**/[?***$parameter***=***$value***][&***$parameter***=***$value***]<br>/page/InC/1.0/<br>/page/InC/1.0/?language=es<br>/page/NoC-CR/1.0/?language=nl&relatedURL=http://example.com/x |

Additional information about the requirements for HTTP interaction behaviour for each of these resource types and URI patterns can be found in the “HTTP interaction patterns” subsection of the “Publication and implementation” section below.

## Use of URIs in metadata and linking

When an adopter refers to a Rights Statement in metadata (e.g. in a set of RDF triples about an item), the URI associated with the Rights Statement **MUST** be used, as presented in the examples later in this document.

This raises issues for representing rights statuses requiring extra information, such as an expiry date or the URL of a contract that further specifies restrictions on the use of an object, in an object's metadata. There are several options available, such as using ODRL's fine-grained permission framework or ccREL's properties, e.g.

*cc:deprecatedOn*.

The issue has been deemed out-of-scope for this document, as it does not impact the specification of the technical infrastructure required for RightsStatements.org. However, we present non-normative examples in the sections “Data Modelling” and “Object Metadata Examples.” Note that these are only *possible* ways to represent such “customized” statements. Further guidance will be provided by implementers of RightsStatements.org like DPLA and Europeana for their data providers, as the standardized Rights Statements become available for them.

Human-readable representation URIs **MAY** be used for linking to specific representations in an application’s public user interface. Human-readable representation URIs **MUST NOT** be used as a substitute for rights statement URIs in provider metadata.

Human-readable representation URIs **MAY** be created dynamically, or stored separately, either for human or API consumption. Human-readable representations **MAY** contain a language parameter that can be used to identify a request for a specific translation. The language tag value used with a `language` parameter **MUST** comply with IETF BCP47.[6]

For example, an aggregator could link to the following human-readable representation URI for an NoC-NC statement in the Spanish version of their site using a URI such as the following. The URI would return a translation of the rights statement text in Spanish.

```
http://rightsstatements.org/page/NoC-NC/1.0/?language=es
```

## Additional metadata

Through the revision process of both the previously released *Recommendations for the Technical Infrastructure for Standardized Rights Statements* and the Rights Statements themselves, additional requirements arose for which certain statements could have an additional metadata element. For example, in the context of the InC-OW-EU statement ("In Copyright - EU Orphan Work"), it can be desirable to include a URI for an associated record in the European Union Orphan Works Database.[7] When required in the context of a metadata aggregator or one of its data providers, this additional metadata will normally take the form of additional RDF statements about the object or the applied rights statement, as shown in object metadata examples later in this document. However, in some cases it might be desirable to link to a human-readable representation that will display this additional information.

The URI for human-readable representations of specific rights statements **MAY** include an additional query string parameter to display additional information. These parameters are *only* valid when used in the URI for a human-readable representation. These parameters **MUST NOT** be used as part of rights statement URIs or in URIs for machine-readable representations. A parameter listed for a given rights statement in the table below **MUST** be used only for the purpose indicated for a given Rights Statement vocabulary term. Parameters for additional metadata **MUST NOT** be used in combination with any Rights Statement not included in the table below; otherwise, it is considered an *invalid* use of that parameter. The behaviour of the interaction patterns between a client and a server in cases of valid and invalid use of parameters is described in the "HTTP interaction patterns" subsection of the "Publication and implementation" section below.

| **Identifier** | **Parameter** | **Purpose/notes** |
|---|---|---|
| InC-OW-EU | `relatedURL` | Associated record in the EU orphan works database |
| NoC-NC | `date` | Contact expiry date; must be an ISO 8601 calendar date |
| NoC-CR | `relatedURL` | URL with more information on the contractual restrictions |
| NoC-OKLR | `relatedURL` | URL with more information on other known legal restrictions |

[6] https://tools.ietf.org/html/bcp47
[7] https://euipo.europa.eu/orphanworks/

# Data Modelling

A goal of this work is to develop a "simple, flexible, and descriptive" framework that allows organizations to communicate the rights status of resources contributed to large-scale aggregations like Europeana or the DPLA. In service of this objective, the group reviewed existing schemas for expressing rights information. Because the Rights Statements are not licences, per se, the group believed that the use of these standards out of the box could lead to confusion among implementers. Therefore, the data modelling efforts have focused on describing and organizing Rights Statements. These considerations led to agreement on some basic principles:

- The working group will model the Rights Statement metadata using the Resource Description Framework (RDF) 1.1 Abstract Syntax[8] as a Simple Knowledge Organization System (SKOS) concept scheme.[9]
- The model will treat Rights Statements as members of the classes `skos:Concept` and `dcterms:RightsStatement`.
- Related Rights Statements will be gathered into a `skos:Collection`.
- The Rights Statements model requires all literals and/or lexical labels to include an appropriate language tag using RFC5646/BCP47.[10]

### Extensibility

Because the rights status of a particular resource might involve a number of additional facets beyond the concepts defined here, the group discussed the feasibility of various extensions to the framework. Current decisions regarding extensibility made by the Technical Working Group include:

- Publishing designations of validity (e.g. expiry dates) for a statement instance has been determined to be out-of-scope for the SKOS concept scheme.
- Instead, recommendations for HTTP interaction patterns have been included, which address the issue of providing human-readable recommendations for some of these extensions.
- Non-normative examples have been included for providers to demonstrate possible ways to express additional data on the status of their objects in the metadata for these objects.
- Recommendations for incorporating additional rights and access related properties to aggregation-specific best practices (say, at the level of Europeana and DPLA) have been deferred.

## Class for Rights Statements

In attempting to define classes for Rights Statements, the group identified an issue in current practice. Within the Europeana context, both the Europeana Data Model and the Europeana

---

[8] http://www.w3.org/TR/2014/REC-rdf11-concepts-20140225/
[9] http://www.w3.org/TR/2009/REC-skos-reference-20090818/
[10] As per http://www.w3.org/TR/rdf11-concepts/#dfn-language-tag

Licensing Framework have adopted the Creative Commons Rights Expression Language (ccREL) `cc:License` class.[11] The CC REL RDF Schema asserts that a `cc:License` is a subclass of `dcterms:LicenseDocument` *(*"A legal document giving official permission to do something with a Resource"*)* which appears narrower than what is intended by the definition of `cc:License` ("a set of requests/permissions to users of a Work, e.g. a copyright licence, the public domain, information for distributors").[12]

Because Rights Statements are not legal documents, this group feels that using `cc:License` may be misleading, especially in cases that express public domain status. Therefore, this version of the Rights Statement concept scheme uses the broader `dcterms:RightsStatement` class ("A statement about the intellectual property rights (IPR) held in or over a Resource, a legal document giving official permission to do something with a resource, or a statement about access rights")[13].

As the statements are part of a SKOS concept scheme that groups them into a controlled whole (subdivided in general SKOS collections), they are also typed as `skos:Concept`.

The basic patterns for our Rights Statements data is thus the following (for clarity, we show here the relation between a cultural object and its rights statement, which we'll come back to later in this document).

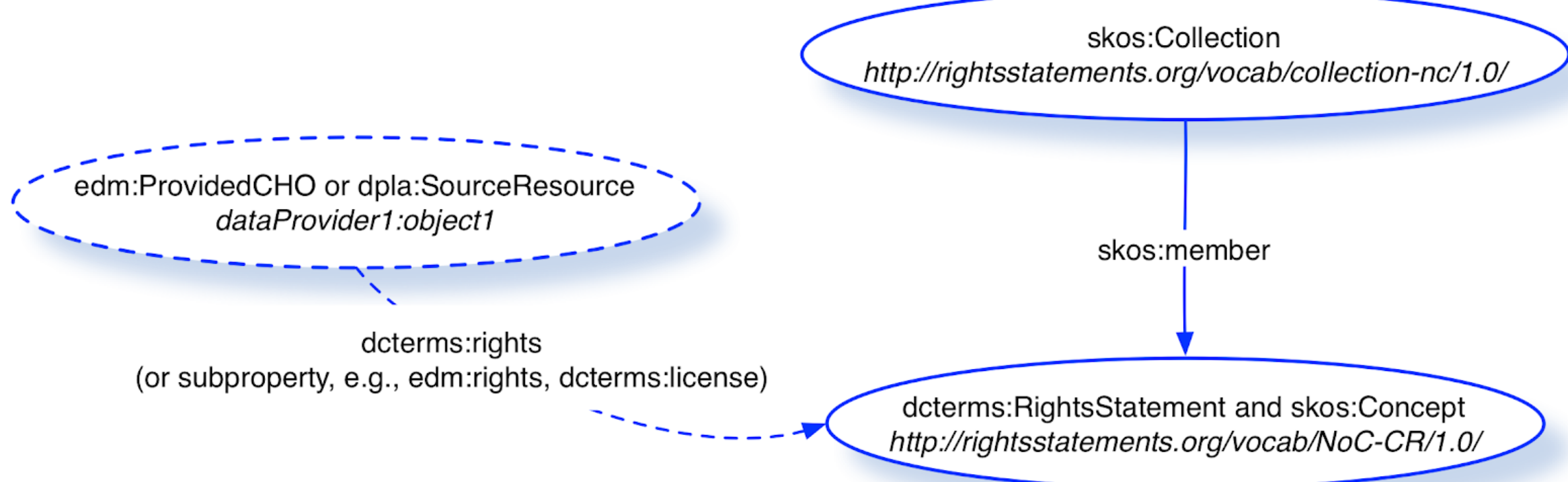

## Interoperability and comparison with other frameworks for rights and licensing

The choices we made for representing our Rights Statements are compatible with a number of the frameworks mentioned above. In particular, our choice for `dcterms:RightsStatement` as the central resource can accommodate most statements expressed according to the recommendations of Dublin Core, ODRS, ccREL and ODRL:

- `dcterms:LicenseDocument` is a subclass of `dcterms:RightsStatement` and thus instances of this class will naturally fit as statements from our perspective

[11] http://www.w3.org/Submission/ccREL/
[12] Further investigation reveals that is this partly due to early CC adoption of `cc:License` prior to the addition of classes to the DCMI Metadata Terms in 2008.
[13] http://purl.org/dc/terms/RightsStatement

- `odrs:License` is a subclass of `dcterms:LicenseDocument` and thus of `dcterms:RightsStatement`
- `cc:License` is a subclass of `dcterms:LicenseDocument`
- `odrl:Policy` does not have formal semantic relationships with above mentioned classes, but instances of `cc:License` have been described as instances of `odrl:Policy` in the ODRL documentation and they can appear as objects of `dcterms:rights` statements about assets being licensed (which then formally makes them instances of the `dcterms:RightsStatement` class).

Further compatibility with the more advanced frameworks shows the way we envision describing rights statements that have additional metadata attached to them. In particular, our approach is compatible with following the ODRL patterns for permissions and the information that a specific statement derives from another one, as in the following example of a custom rights statement with an explicit expiry date:[14]

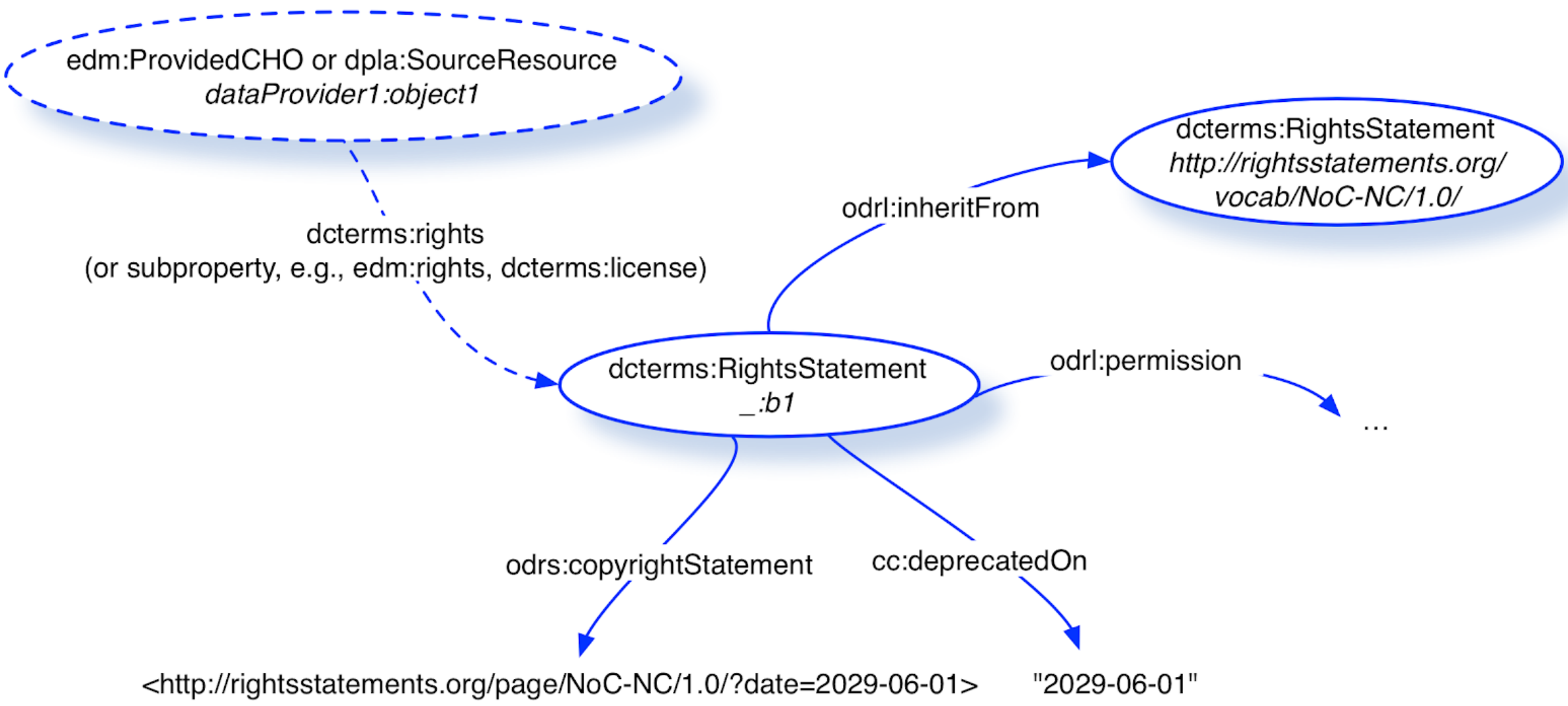

This diagram also shows the parallel between our approach and ODRS' "dual-resource" pattern as illustrated in the ODRS documentation:[15]

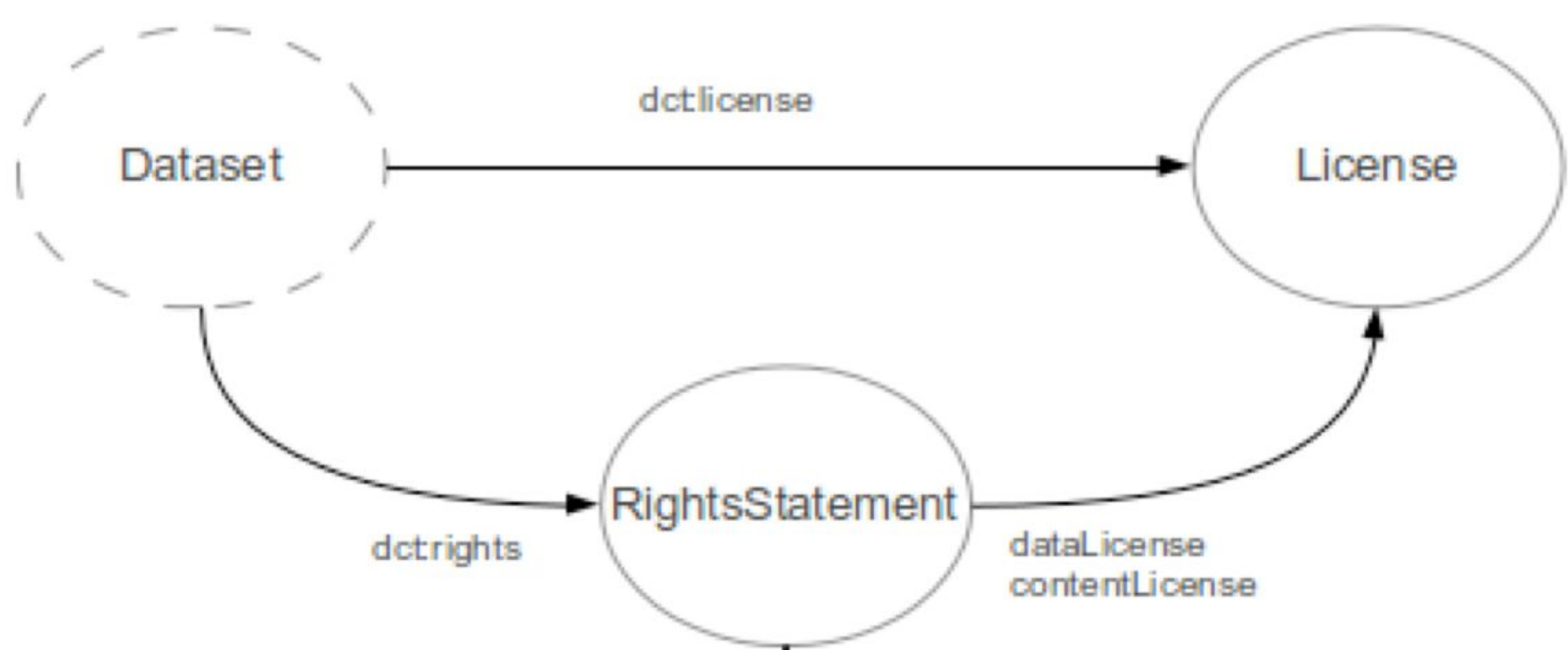

[14] Note that in accordance with the current situation about expressing extra metadata on the rights statements applicable for objects, this example presents only a *possible* way to represent the situation at hand. Europeana and the DPLA will provide further guidance about this issue.
[15] http://schema.theodi.org/odrs/

Our pattern is very similar, with the exception of the `dct:license` link between the original resource and the base statement/licence, and the property we use to relate the base statement to the derived one (we re-use an ODRL property).

Please keep in mind that this corresponds to an advanced (and yet not completely specified) use of the framework, where Rights Statements are "customized" for specific resources. The needs of many data providers will be covered by the "basic" rights statements made available by our service or Creative Commons. For these cases, there is no need for the more complex ODRS pattern.

## Domain of RightsStatements

Although RightsStatements.org was initially developed in the context of large-scale cultural heritage aggregators such as Europeana and DPLA, we recognize that there is interest in their use beyond these specific implementations.[16] As noted above, Rights Statements are represented as SKOS vocabulary and declared as members of the *`dcterms:RightsStatement`* class. In this model, a rights statement must apply broadly to *resources* defined by the DCMI Abstract Model and the Resource Description Framework.[17] The requirements defined in these recommendations should allow the use of Rights Statements in any RDF conformant schema.

However, the use of a Rights Statement may be constrained to narrower classes within a particular application, as demonstrated by the Europeana Data Model.[18] The EDM example should not prevent other interpretations that assign rights statements to other valid classes of resources (see Wikidata example below).

## Property for Rights Statement Labels

Each Rights Statement will have a primary human-readable label ("In Copyright") and short identifier ("InC"). Related standards for expressing copyright status (CC, EDM) use *`dc:title`* while ODRS uses *`rdfs:label`*. In this version of the RightsStatements.org concept scheme, we propose using *`skos:prefLabel`* in line with our secondary goal of creating a SKOS vocabulary for the Rights Statements.

## Community Specific Permissions & Constraints

A feature of the proposed Rights Statements includes community-specific permissions and constraints, for example "In Copyright - Educational Use Only." Using the Open Digital Rights Language (ODRL) Version 2.1 Ontology[19], we propose using *`odrl:permission`* with *`odrl:purpose`* specifying "educational." As we could not identify an external vocabulary supporting "educational use" for this scheme, it would necessitate hosting and creating a term, e.g., `http://rightsstatements.org/vocab/educationalUse`. This larger question of hosting and maintaining community-specific constraints in addition to Rights Statements at

[16] Urban, R.J. (2019) What are RightsStatements About? *RightsStatements.org* https://rightsstatements.org/en/2019/05/what-are-rights-statements-about.html
[17] https://www.dublincore.org/specifications/dublin-core/abstract-model/2007-06-04/
https://www.w3.org/TR/rdf-schema/
[18] https://pro.europeana.eu/resources/standardization-tools/edm-documentation
[19] http://www.w3.org/ns/odrl/2/

RightsStatements.org requires further discussion. This issue has thus been postponed, and we welcome feedback from the community about it.

### Rights Statements as Linked Data

The Rights Statements are most valuable as Linked Data if they enable connections to other existing frameworks for expressing rights information. Whenever possible, a RightsStatements.org RDF representation includes references to related standards through the use of *skos:closeMatch*, *skos:exactMatch*, *skos:broadMatch*, *skos:narrowMatch*, or *skos:relatedMatch*. For example, the PREMIS data model allows for the inclusion of a small set of coded rights status statements.[20] The RightsStatements.org data can reflect this by including the following assertion:

```
<http://rightsstatements.org/vocab/InC/1.0/>
      skos:relatedMatch premiscopy:cpr .
```

The following figure also shows possible links between the RightsStatements.org "In Copyright" statement and the current Europeana statements related to it:

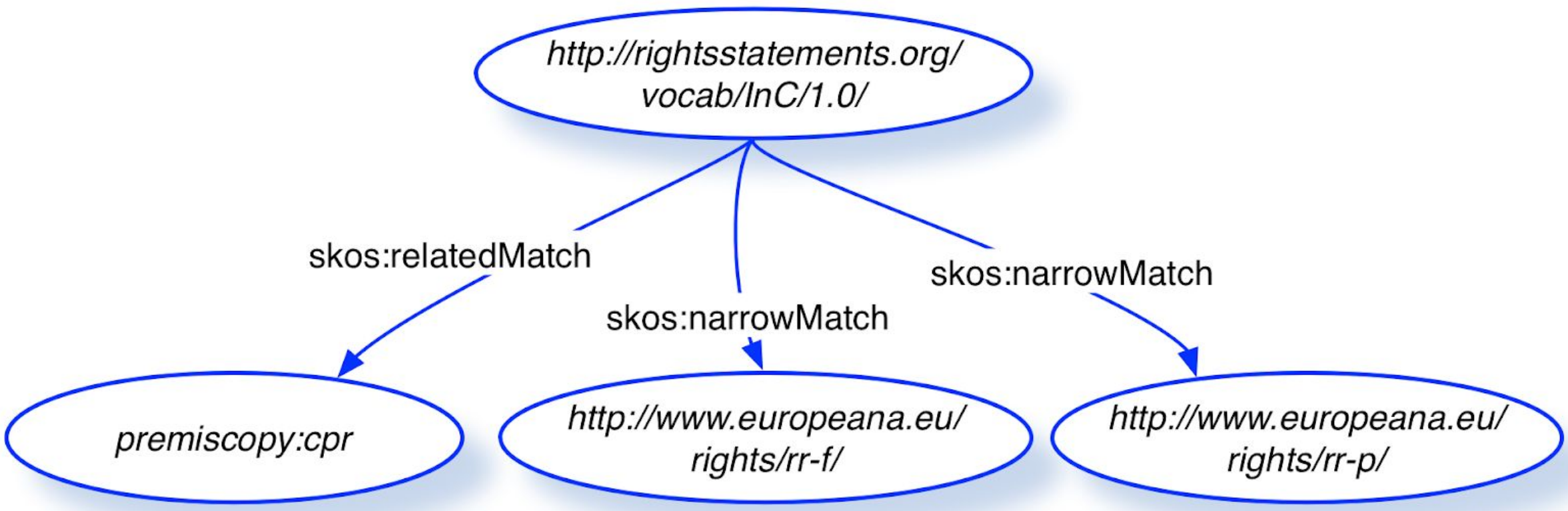

As the RightsStatements.org concept scheme develops, we will seek to incorporate relationships to other rights expressions and classifications deemed appropriate for the cultural heritage community.

## Technical Editorial Policies

### Changes

Changes to Rights Statements are governed by the RightsStatements.org Editorial Policy. In principle, there are three different types of changes that can occur:

- The addition or removal of one or more Rights Statement(s) to the existing set of Rights Statements.

[20] http://www.loc.gov/standards/premis/

- Minor changes to the literal values of one or more existing Rights Statements that do not alter the semantic meaning of the statements
- Significant changes to the literal values of one or more existing Rights Statement that alter the semantic meaning of the statements.

## Versions

Published versions of the Rights Statements are static in meaning. To support changes through a periodic editorial process, statements **MUST** have a version number. The initial version of the original set of Rights Statements is published as version 1.0. Statements introduced later are published at the version number that matches the version number of the vocabulary that is current at the time of addition (see following). All Rights Statements **MUST** move to a higher version number at the same time.

Each statement **MUST** belong to a concept scheme, representing a version of the vocabulary, which shares a version number with its members. Concept schemes **MUST** contain a version of each statement still recommended for use at the time of its publication. The effect is that changes to statements will only take place within the context of an update to the broader scheme.

The version of a statement **MUST** be canonically given by its *owl:versionInfo* property. Additionally, the version is included in the URI for the statement as described in “URI Design”, above. Version numbers are not considered meaningful, except insofar as successive version numbers **MUST** be greater than those of their predecessors.

## Addition or removal of Rights Statements

The addition or removal of Rights Statements is governed by the RightsStatements.org Editorial policies. Once a change is approved, the technical working will apply changes to the representation of record. An appropriate comment documenting the changes **MUST** be included for all committed changes.

The addition of new statements does not require a new version. For new statements being added to an existing version, a new URI using the current URI version pattern **MUST** be created for the new statement. This URI **MUST** be added to the concept scheme.

Deprecated Rights Statements **MUST** only be removed from new versions of the vocabulary. If deprecated statements are replaced by statements in the new version, a *dcterms:isReplacedBy* triple **SHOULD** be added to to the superseded version in order to redirect users to the preferred statement in the new version. If desired, a *skos:historyNote* **MAY** be added to indicate the reasons why the statement was deprecated, with review according to the Editorial Policy.

## Minor changes

Minor changes to the human-readable text (literal values) of an existing Rights Statement do not require a new version. Changes can be considered to be minor if they do not affect the semantic meaning of the statement(s). Given that, minor changes will be limited to corrections of spelling mistakes and other errors. Minor changes to existing Rights Statements will be managed by the International Rights Statements Working Group in accordance with the RightsStatements.org Editorial Policy. Once a change is approved, the Technical Working Group will apply it to the Turtle-serialized SKOS and push it to the version control system. An appropriate comment documenting the changes **MUST** be included for all committed changes.

## Substantive changes

Substantive changes in human-readable text require a new version of the Rights Statement. In this context, substantial changes **MUST** be understood as changes in wording or structure of the statements that alter the semantics of the Rights Statements. The substantive meaning of a statement should never be altered. If there is a need to change the substance of a statement then the existing statement **MUST** be deprecated and replaced by a new statement. Versioning of the Rights Statements **MUST** happen for all non-deprecated statements at the same time. Substantive changes to existing Rights Statements will be managed by the International Rights Statements Working Group in accordance with the Rightsstatements.org Editorial Policy. Once a change is approved, the Technical Working Group will implement the new version on the staging server for review. Once the new versions are approved, the Turtle-serialized SKOS will be pushed to the version control system. An appropriate comment documenting the changes **MUST** be included for all committed changes.

Subsequent versions using the same statement identifier **MAY** be published; these are considered to be distinct entities, represented by unique URIs and supersede all prior versions. When a new version is published, a triple **MUST** be added to machine-readable representations of the previous versions relating it to the new version with *`dcterms:isReplacedBy`*. The inverse relationship (*`dcterms:replaces`*) **SHOULD** be included in machine-readable representations of subsequent statement versions

## Translations

The procedure for approving and adding new translations of Rights Statements is managed by the Rightsstatements.org Editorial Policy. It is assumed that the RightsStatements.org technical infrastructure can handle an unlimited number of languages provided that the translations conform to the following requirements:

- Human-readable representations MAY contain a language parameter that can be used to identify a request for a specific translation. The language tag value used with a language parameter MUST comply with IETF BCP47.[21]
- Human-readable translations MUST conform to the implementation guidelines specified elsewhere in this document.

Once a translation is approved, the Technical Working Group will integrate the translation with the current published version of the Turtle-serialized SKOS on the staging server for review. Once RightsStatements.org and the translation partner have approved the staged translation, the technical working group will push the changes to the production website. Credit for the translation team will be added to the RightsStatements.org acknowledgements page and as a `skos:editorialNote` attached to the concept scheme.

[21] https://tools.ietf.org/html/bcp47

# Human and Machine readability

The Rights Statements vocabulary will contain a human and machine-readable overview. Additionally, each Rights Statement will be available in human and machine-readable versions. The human-readable version will be rendered in HTML generated by the RDF serializations. This section deals with the response a machine gets when a Rights Statement is requested. RightsStatements.org **MUST** offer the following formats:

- HTML5 with RDF(a)
- JSON-LD
- Turtle RDF syntax

The last two will be accessible through content negotiation using HTTP requests, following the recipes presented in the section "Publication and Implementation".

Machine-readable representations are used to structurally communicate information about the Rights Statements following the data model described above. Human-readable representations on RightsStatements.org are intended to present, at a minimum, the statement's title, descriptive and scope information, jurisdiction, creator, version, and other translations.

Properties and classes will be drawn from the following sources:

| Source | Prefix abbreviation | Namespace |
|---|---|---|
| Creative Commons Rights Expression Language (ccREL) | `cc:` | `http://creativecommons.org/ns#` |
| Dublin Core Elements 1.1 | `dc:` | `http://purl.org/dc/elements/1.1/` |
| DCMI Type Vocabulary | `dcmitype:` | `http://purl.org/dc/dcmitype/` |
| DCMI Metadata Terms | `dcterms:` | `http://purl.org/dc/terms/` |
| Europeana Data Model | `edm:` | `http://www.europeana.eu/schemas/edm/` |
| ODRL | `odrl:` | `http://www.w3c.org/ns/odrl/2/` |
| PREMIS Copyright Status | `premiscopy:` | `http://id.loc.gov/vocabulary/preservation/copyrightStatus/` |
| SKOS | `skos:` | `http://www.w3.org/2004/02/skos/core#` |
| OWL | `owl:` | `http://www.w3.org/2002/07/owl#` |
| ODRS | `odrs:` | `http://schema.theodi.org/odrs#` |
| Wikidata Entities | `wd:` | `http://www.wikidata.org/entity/` |
| Wikidata Properties | `wdt:` | `http://www.wikidata.org/prop/direct/` |

### Example Rights Statement in Turtle (RDF syntax)

The following example demonstrates the “In Copyright - Educational Use Permitted” Rights Statement expressed in RDF. This articulation is meant to illustrate how the Rights Statements might be expressed and is not a definitive version of the statement or properties applied.[22]

```
@prefix dc: <http://purl.org/dc/elements/1.1/> .
@prefix dcmitype: <http://purl.org/dc/dcmitype/> .
@prefix dcterms: <http://purl.org/dc/terms/> .
@prefix owl: <http://www.w3.org/2002/07/owl#> .
@prefix premiscopy: <http://id.loc.gov/vocabulary/preservation/copyrightStatus/>
.
@prefix skos: <http://www.w3.org/2004/02/skos/core#> .

<http://rightsstatements.org/vocab/InC-EDU/1.0/>
     a dcterms:RightsStatement, skos:Concept ;
  skos:prefLabel "In Copyright - Educational Use Permitted"@en ;
  dcterms:modified "2015-10-16"^^xsd:date ;
  skos:definition """This Item is protected by copyright and/or related rights.

     You are free to use this Item in any way that is permitted by the copyright
     and related rights legislation that applies to your use. In addition, no
     permission is required from the rights-holder(s) for educational uses.

     For other uses, you need to obtain permission from the
     rights-holder(s)."""@en ;
  skos:note "Unless expressly stated otherwise, the organization that has made
     this Item available makes no warranties about the Item and cannot guarantee
     the accuracy of this Rights Statement. You are responsible for your own
     use."@en ;
  skos:note "You may need to obtain other permissions for your intended use. For
     example, other rights such as publicity, privacy or moral rights may limit
     how you may use the material."@en ;
  skos:note "You may find additional information about the copyright status of
     the Item on the website of the organization that has made the Item
     available."@en ;
  skos:scopeNote "This Rights Statement can be used only for copyrighted Items
     for which the organization making the Item available is the rights-holder
     or has been explicitly authorized by the rights-holder(s) to allow third
     parties to use their Work(s) for educational purposes without first
     obtaining permission."@en ;
  skos:relatedMatch premiscopy:cpr ;
  dcterms:creator <http://rightsstatements.org/vocab/irswg> ;
  owl:versionInfo "1.0" ;
  dc:identifier "InC-EDU" ;
  skos:inScheme <http://rightsstatements.org/vocab/1.0/> .
```

# Object Metadata Examples

This section outlines how an object can incorporate Rights Statements in its metadata. All examples in this section are understood to be non-normative. In this section we use the following namespace abbreviation prefixes (in Turtle syntax):

[22] The most up-to-date version of the rights statements as expressed in Turtle-serialized SKOS can be found at https://github.com/rightsstatements/data-model.

```
@prefix cc: <http://creativecommons.org/ns#> .
@prefix dc: <http://purl.org/dc/elements/1.1/> .
@prefix dcterms: <http://purl.org/dc/terms/> .
@prefix edm: <http://www.europeana.eu/schemas/edm/> .
@prefix dpla: <http://dp.la/about/map/> .
@prefix ore: <http://www.openarchives.org/ore/terms/> .
@prefix rdfs: <http://www.w3.org/2000/01/rdf-schema#> .
@prefix skos: <http://www.w3.org/2004/02/skos/core#> .
@prefix foaf: <http://xmlns.com/foaf/0.1/> .
```

## Objects Available at Europeana

The following example describes rights information for the object "Stanton Harcourt, Church"[23]. Prior to the availability of Rights Statements, it was assigned the Europeana-minted statement "Rights Reserved - Free Access".[24] Now the British Library and Europeana can use the "In Copyright" Rights Statement.

```
<http://data.europeana.eu/aggregation/provider/92037/_http___www_bl_uk_onlinegal
lery_onlineex_topdrawings_s_zoomify85637_html> a ore:Aggregation ;
     edm:aggregatedCHO
<http://data.europeana.eu/item/92037/_http___www_bl_uk_onlinegallery_onlineex_to
pdrawings_s_zoomify85637_html> ;
     edm:dataProvider "The British Library" ;
     edm:isShownAt
<http://www.bl.uk/onlinegallery/onlineex/topdrawings/s/zoomify85637.html> ;
     edm:provider "The European Library"@en ;
     edm:rights <http://rightsstatements.org/vocab/InC/1.0/> .

<http://data.europeana.eu/item/92037/_http___www_bl_uk_onlinegallery_onlineex_to
pdrawings_s_zoomify85637_html> a edm:ProvidedCHO ;
     dc:title "Stanton Harcourt, Church" ;
     dc:creator "Artist : Grimm, Samuel Hieronymus" ;
     dc:type "manuscript" ;
     dc:description "The 12th-century Church of St Michael contains the tomb of
Robert Harcourt, Henry VIII's standard bearer at the Battle of Bosworth,
1485"@en ;
     dc:rights "Copyright © British Library Board"@en .
```

The following example presents a *possible* way to describe rights information for the object "La polka, vaudeville en un acte"[25], previously assigned the Europeana-minted statement "Out of copyright - non commercial re-use"[26] with an expiry date set to 17 November 2029. One can now use the "No Copyright - Non Commercial Use Only" Rights Statement:

```
<http://data.europeana.eu/aggregation/provider/9200332/BibliographicResource_300
0123583360> a ore:Aggregation ;
```

[23] http://www.europeana.eu/portal/record/92037/_http___www_bl_uk_onlinegallery_onlineex_topdrawings_s_zoomify85637_html.html

[24] http://www.europeana.eu/rights/rr-f/

[25] http://www.europeana.eu/portal/record/9200332/BibliographicResource_3000123583360.html

[26] http://www.europeana.eu/portal/rights/out-of-copyright-non-commercial.html

```
     edm:aggregatedCHO
<http://data.europeana.eu/item/9200332/BibliographicResource_3000123583360> ;
     edm:dataProvider "Österreichische Nationalbibliothek - Austrian National
Library" ;
     edm:isShownAt
<http://digital.onb.ac.at/OnbViewer/viewer.faces?doc=ABO_%2BZ175802906> ;
     edm:provider "The European Library"@en ;
     edm:rights [
      odrl:inheritFrom <http://rightsstatements.org/vocab/NoC-NC/1.0/> ;
      cc:deprecatedOn "2029-11-17" ;
      odrs:copyrightStatement
      <http://rightsstatements.org/page/NoC-NC/1.0/?date=2029-11-17>
      ]

<http://data.europeana.eu/item/9200332/BibliographicResource_3000123583360> a
edm:ProvidedCHO ;
     dc:title "La polka, vaudeville en un acte" ;
     dc:creator "Guinot, Eugene", "Berat, Frederic" .
```

## Object Available at DPLA

The following example presents a *possible* way to describe rights information for the object "Educational institution study":[27]

```
<http://dp.la/api/items/fc69709e798f9ad881cf302953ad4c83> a ore:Aggregation ;
      edm:aggregatedCHO
<http://dp.la/api/items/fc69709e798f9ad881cf302953ad4c83#sourceResource> ;
      edm:rights <http://rightsstatements.org/vocab/InC-EDU/1.0/> .

<http://dp.la/api/items/fc69709e798f9ad881cf302953ad4c83#sourceResource>
      a dpla:SourceResource ;
      dc:rights "Access to the Internet Archive's Collections is granted for
scholarship and research purposes only. Some of the content available through
the Archive may be governed by local, national, and/or international laws and
regulations, and your use of such content is solely at your own risk" ;
      dc:creator "Boston Redevelopment Authority" ;
      dc:title "Educational institution study" .
```

## Object Available at DPLA, expressed in Schema.org

The following is an example of an object from with its rights information adapted from the object “Moore Manufacturing Company” available at DPLA:[28]

```
{
      "@context": "https://schema.org/",
      "@id": "https://dp.la/item/00029b55b2cff2455d8902d01f51159f",
      "@type": "Photograph",
      "name": "Moore Manufacturing Company",
      "url": "https://dmr.bsu.edu/digital/collection/sellers/id/389",
      "license": "http://rightsstatements.org/vocab/NoC-US/1.0/",
```

[27] http://dp.la/item/fc69709e798f9ad881cf302953ad4c83
[28] https://dp.la/item/00029b55b2cff2455d8902d01f51159f

```
      "provider": [
            {
             "@type": "Organization",
             "name": "Indiana Memory"
            },
            {
             "@type": "Organization",
             "name": "Ball State University. University Libraries. Archives and
Special Collections"
            }],
      "thumbnailURL": "https://dp.la/thumb/00029b55b2cff2455d8902d01f51159f",
      "image": {
            "@type": "ImageObject",
            "url":
                  "https://dmr.bsu.edu/digital/api/singleitem/image/sellers/389/
            default.jpg",
            "license": "http://rightsstatements.org/vocab/NoC-US/1.0/"
      }
}
```

## Object in Local Implementation

The following is an example from the University of California San Diego Library:

```
@prefix lcnaf: <http://id.loc.gov/authorities/names/> .
@prefix premis: <http://www.loc.gov/premis/rdf/v1#> .
@prefix ucsd: <http://library.ucsd.edu/ontology/dams4.2#> .

<obj> dcterms:rights <http://rightsstatements.org/vocab/InC-EDU/1.0/>;
      premis:hasCopyrightJurisdiction "us";
      dcterms:accessRights ucsd:restrictedCampus;
      dcterms:rightsHolder lcnaf:n00085230 .
lcnaf:n00085230 skos:prefLabel "Doe, John, -1993" .
```

## Object Available from Wikidata

The following is an example from Wikidata:

```
@prefix wdt: <http://www.wikidata.org/prop/direct/> .
@prefix wd: <http://www.wikidata.org/entity/> .

<wd:Q50086916> a wikibase:Item ;
    rdfs:label "Aspects of Negro Life"@en ;
    wdt:P31 wd:Q15727816 ; # instance-of: "painting series"
    wdt:P571 "1934-01-01T00:00:00Z"^^xsd:dateTime ; # inception date: 1934
    wdt:P276 wd:Q1060566 ; # location: Schomburg Center for Research in Black
Culture
    wdt:P170 wd:Q4661979 ; # creator: Aaron Douglas
    wdt:P88 wd:Q7257674 ; # commissioned by the Public Works of Art Project
    wdt:P186 wd:Q4259259, wd:Q296955 ; # material used: canvas, oil paint
    wdt:P856 <http://exhibitions.nypl.org/treasures/items/show/170> # official
website
    wdt:P6426 wd:Q47530911 . # RightsStatement status according to source
website
```

```
<wd:Q47530911> a wikibase:Item ;
    rdfs:label "No Copyright - United States"@en ;
    wdt:P31 <wd:Q47530706> ; # instance-of RightsStatements.org statement
    wdt:P2888 <http://rightsstatements.org/vocab/NoC-US/1.0/> # exact match in
the RightsStatements.org vocabulary
    schema:description "RightsStatement.org statement pertaining to works that
are in the Public Domain in the jurisdiction of the United State of America"@en
.
```

Comments in the Turtle syntax (following the # sign) have been added for ease of understanding of the Wikidata properties and values.
Note that Wikidata creates "copies" of RightsStatements.org statements for use in its namespace. These resources are understood to have the exact same semantics as the original one, as represented via the wdt:P2888 statement in the example.

# Publication and Implementation

This section describes the implementation for publishing the Rights Statements in both human and machine-readable forms. Our specification follows the *Best Practice Recipes for Publishing RDF Vocabularies,*[29] and addresses our requirements to provide access to these representations through content negotiation. Our choice of a specific recipe is informed by our need to satisfy all of the minimal and extended requirements as expressed in the *Best Practice Recipes:*[30]

- *M1. The 'authoritative' RDF description of a vocabulary, class, or property denoted by an HTTP URI can be obtained by dereferencing the URI of that vocabulary, class, or property.*
- *M2. The behavior of an HTTP URI denoting an RDFS/OWL vocabulary, class or property, does not lead to inconsistency in the interpretation of the nature of the denoted resource.*
- *E1. 'Human-readable' documentation about an RDF vocabulary, class or property, denoted by an HTTP URI, can be obtained by dereferencing the URI of that vocabulary, class or property.*
- *E2. Applications are able to differentiate between 'versions' of a vocabulary.*

In addition, we introduce three sub-requirements to requirement E1:

- *E1.1. A default translation of 'human-readable' documentation about an RDF vocabulary, class or property, denoted by an HTTP URI, can be obtained by dereferencing the URI of that vocabulary, class or property.*
- *E1.2. Additional translations of 'human-readable' documentation about an RDF vocabulary, class, or property, denoted by an HTTP URI, can be obtained by dereferencing the URI of that vocabulary, class, or property, along with the inclusion of an Accept-Language header in the request.*

[29] http://www.w3.org/TR/swbp-vocab-pub/
[30] http://www.w3.org/TR/swbp-vocab-pub/#requirements

- *E1.3. Additional translations are also dereferencable through a common URI pattern when the translation is known to exist.*

As such, the implementation **MUST** follow recipe 6 from the *Best Practice Recipes* ("Extended configuration for a 'slash namespace', using multiple HTML documents and a query service").[31]

## HTTP interaction patterns

In addition to the HTTP interaction patterns specified by *Best Practice Recipes* recipe 6, which specify the necessary content negotiation behavior,[32] we have a few additional requirements that relate to the needs identified above.

*Support for translations.* The implementation **MUST** allow for translations to be served. Additional requirement E1.1 is satisfied by identifying the default translation (English) to return in the server's configuration directives. Translations **MUST** be accessible using content negotiation as specified in E1.2 and through the use of an additional URI component to support requirement E1.3.

*Invalid requests.* When a client makes an invalid request, e.g. when requesting a Rights Statement URI with a query string parameter for additional metadata, the server **MUST** respond to the client accordingly. This is in part by design to ensure that the Rights Statement URIs are applied appropriately. The server **MUST** return a response with an HTTP `406 Not Acceptable`[33] status for invalid requests. If a client requests a Rights Statement URI with query parameters containing an additional metadata payload, an `Alternates` header[34, 35] **SHOULD** be returned with the HTTP `406 Not Acceptable` status. The `Alternates` header in this context functions as a means to provide guidance to a client when an unsuitable representation is requested. A full example follows below:

```
> GET http://rightsstatements.org/vocab/NoC-NC/1.0/?date=2028-01-01
---
< 406 Not Acceptable
< Alternates: {"/page/NoC-NC/1.0/?date=2028-01-01" 0.9 {type text/html}},
{"/vocab/NoC-NC/1.0/" 0.9}
```

*Referencing specific representations.* The `Content-Location` header **SHOULD** be used to allow a generic machine-readable representation URI to refer to the URI for a specific machine-readable representation. URIs for each specific machine-readable representation **SHOULD** be linked from the human-readable representation (e.g. links to download Turtle-serialized RDF can be obtained using the pattern `.../data/x.x/statement.ttl`). In addition, the `Link` header[36] **SHOULD** be used to provide additional link relations[37] between representation URIs and Rights Statement URIs. The `describedby` relation[38] **SHOULD** be returned in a `Link` header for a request for a Rights Statement URI to assert that it is described by the associated readable representation URI. The `derivedfrom` relation[39] in a `Link` header for a request for a representation URI **SHOULD** be used to assert that the specified representation is derived from another representation.

[31] http://www.w3.org/TR/swbp-vocab-pub/#recipe6
[32] http://www.w3.org/TR/swbp-vocab-pub/#negotiation
[33] https://tools.ietf.org/html/rfc7231
[34] https://tools.ietf.org/html/rfc2295
[35] https://tools.ietf.org/html/rfc2296
[36] https://tools.ietf.org/html/rfc5988
[37] http://www.iana.org/assignments/link-relations/
[38] http://www.w3.org/TR/ldp/#link-relation-describedby

[39] https://www.ietf.org/id/draft-hoffman-xml2rfc-21.txt

## Detailed HTTP interaction pattern examples

Dereference the vocabulary URI, requesting HTML content in a specific language:

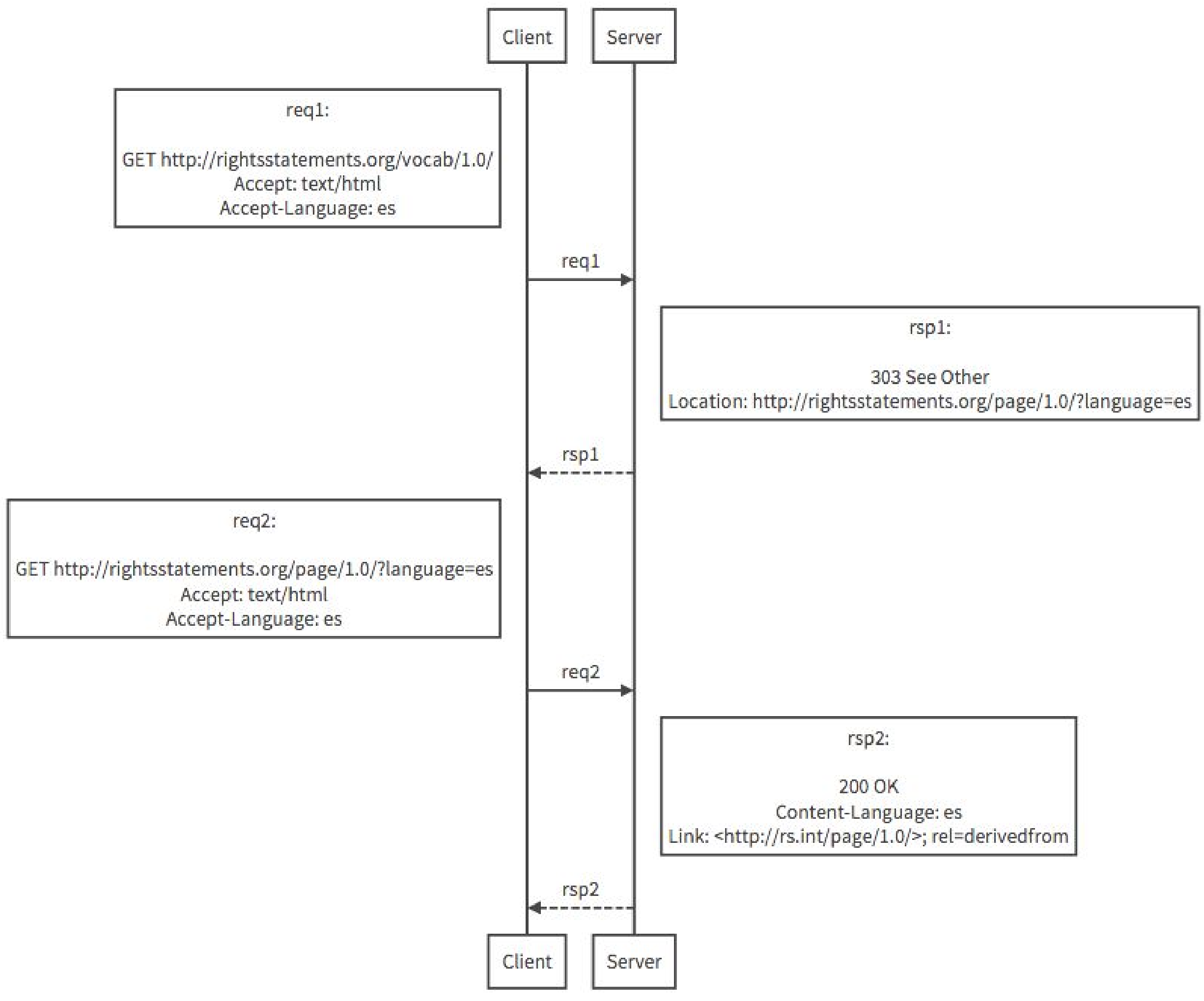

Dereference the URI of a class or property, requesting HTML content in a specific language:

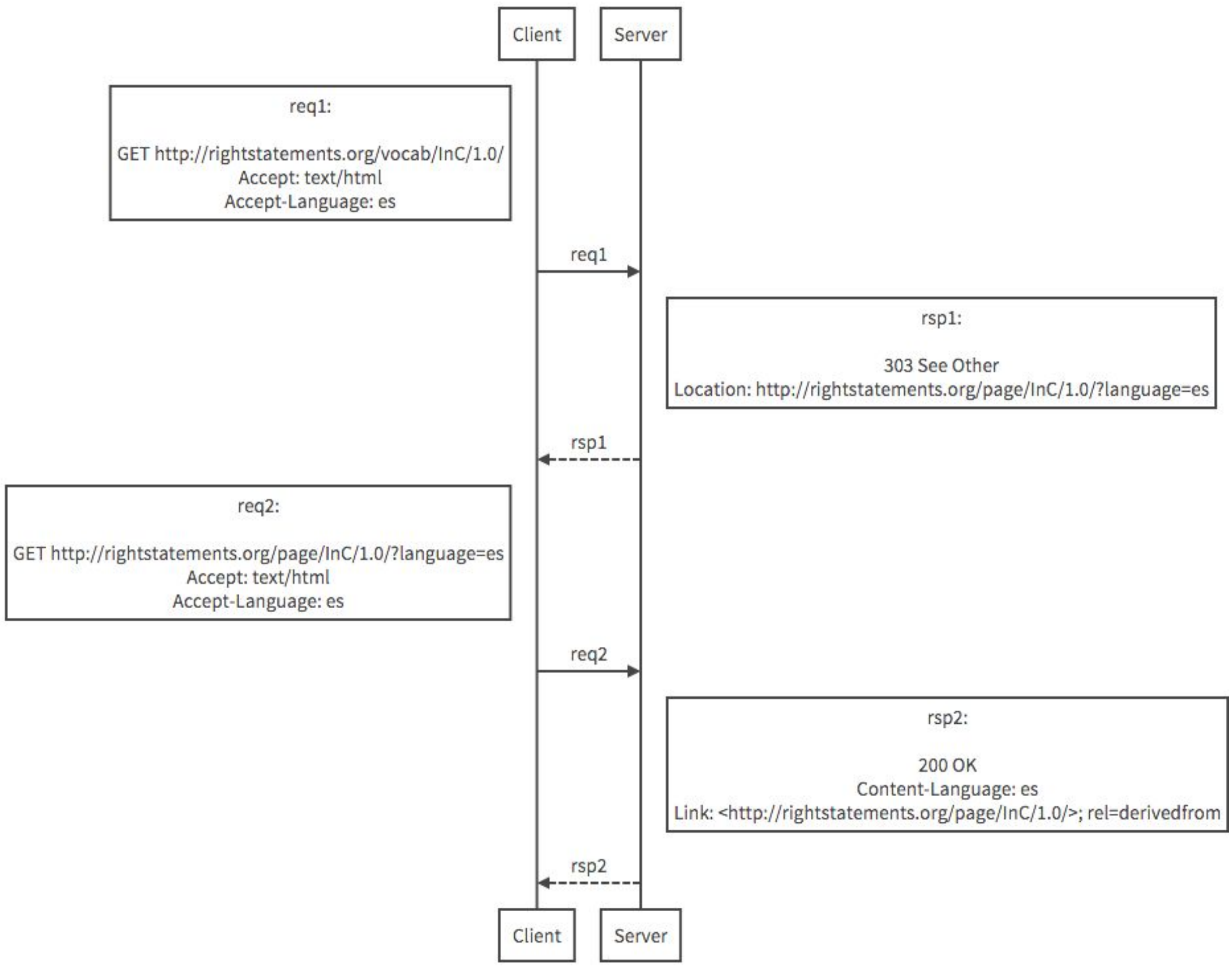

Dereference the URI of a Rights Statement, requesting RDF content:

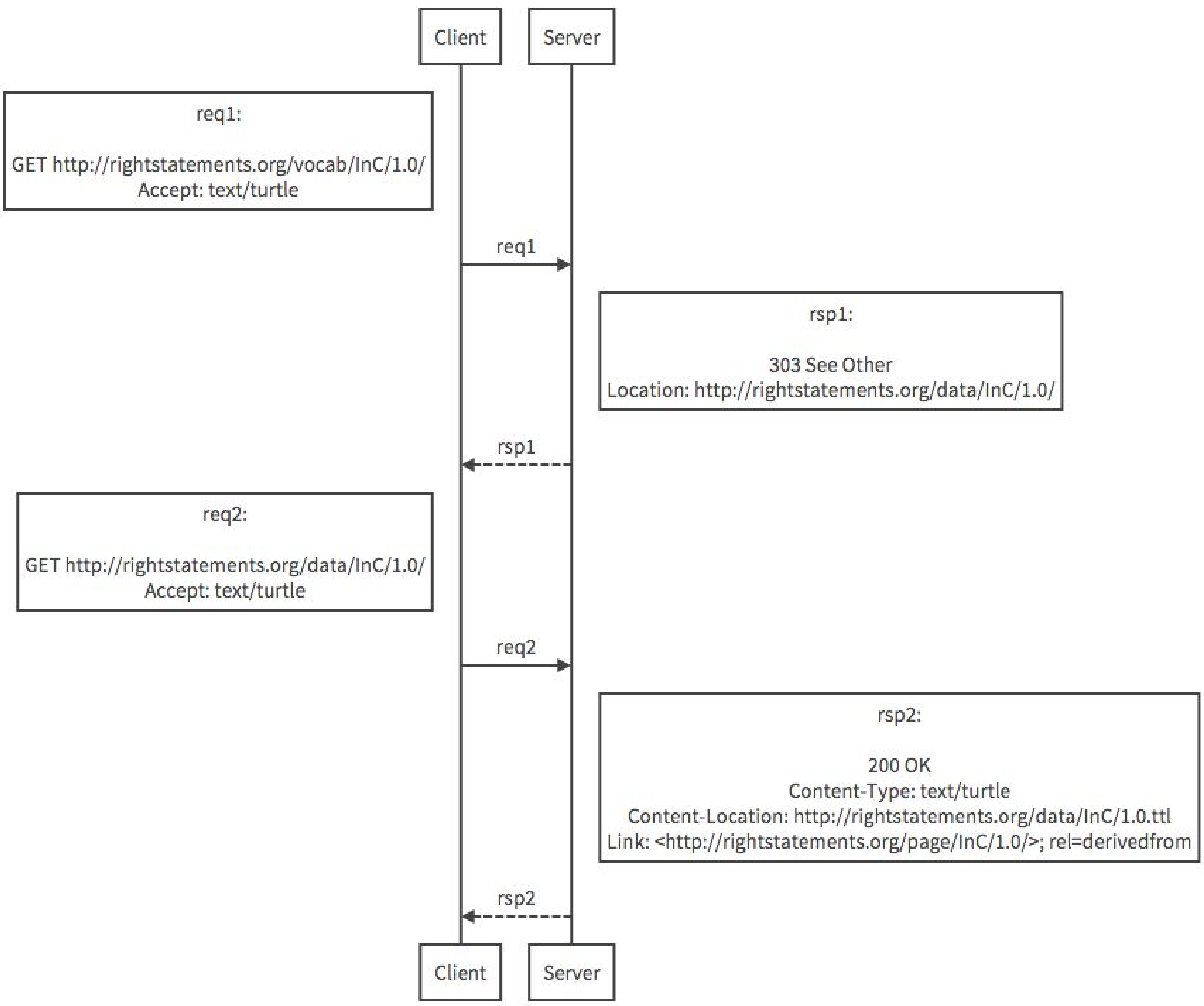

Dereference an NoC-NC statement with expiration date by the human-readable representation URI:

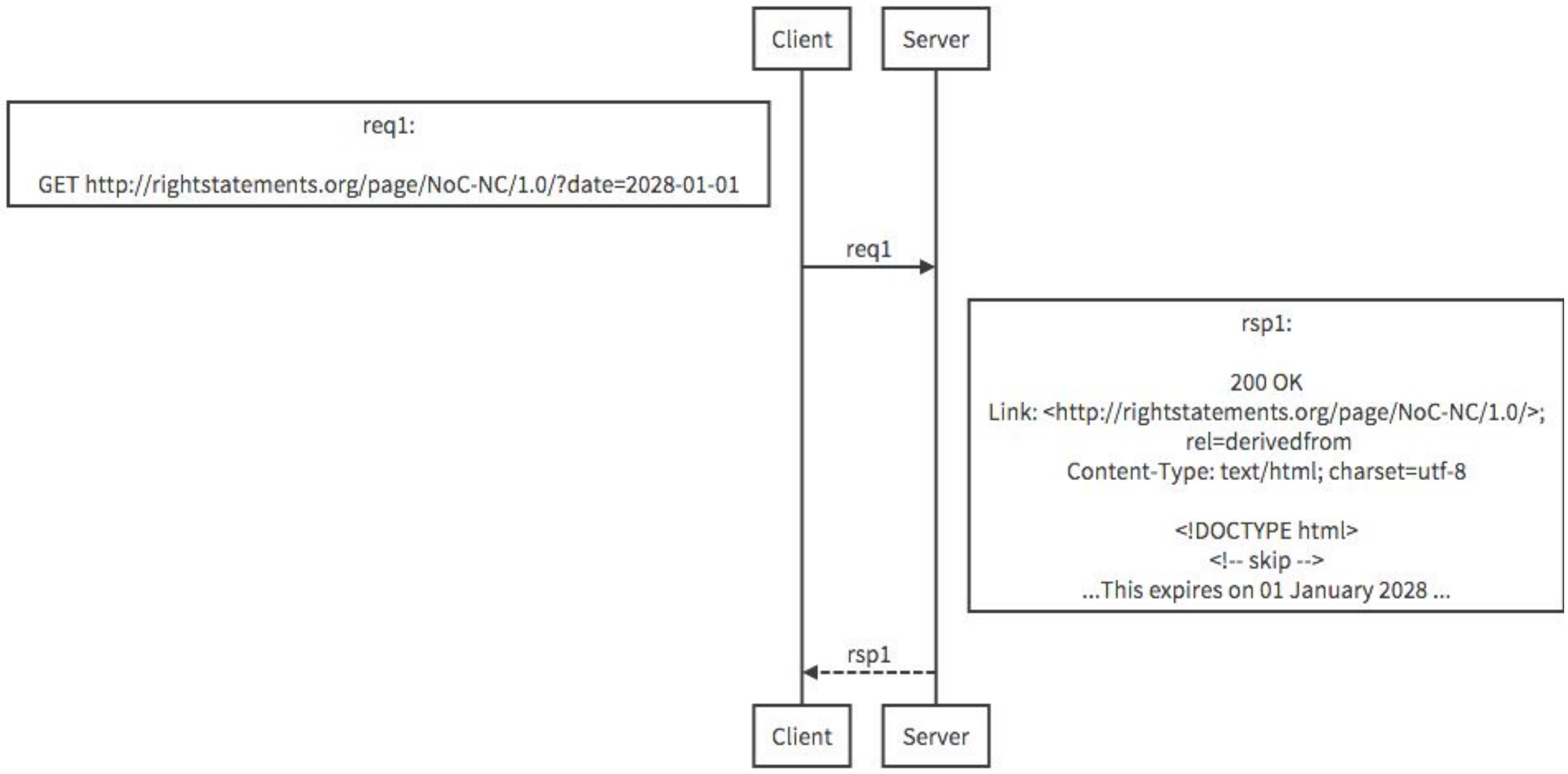

Dereference an NoC-NC statement with expiration date by the Rights Statement URI, requesting RDF (invalid example, with “recovery” guidance):

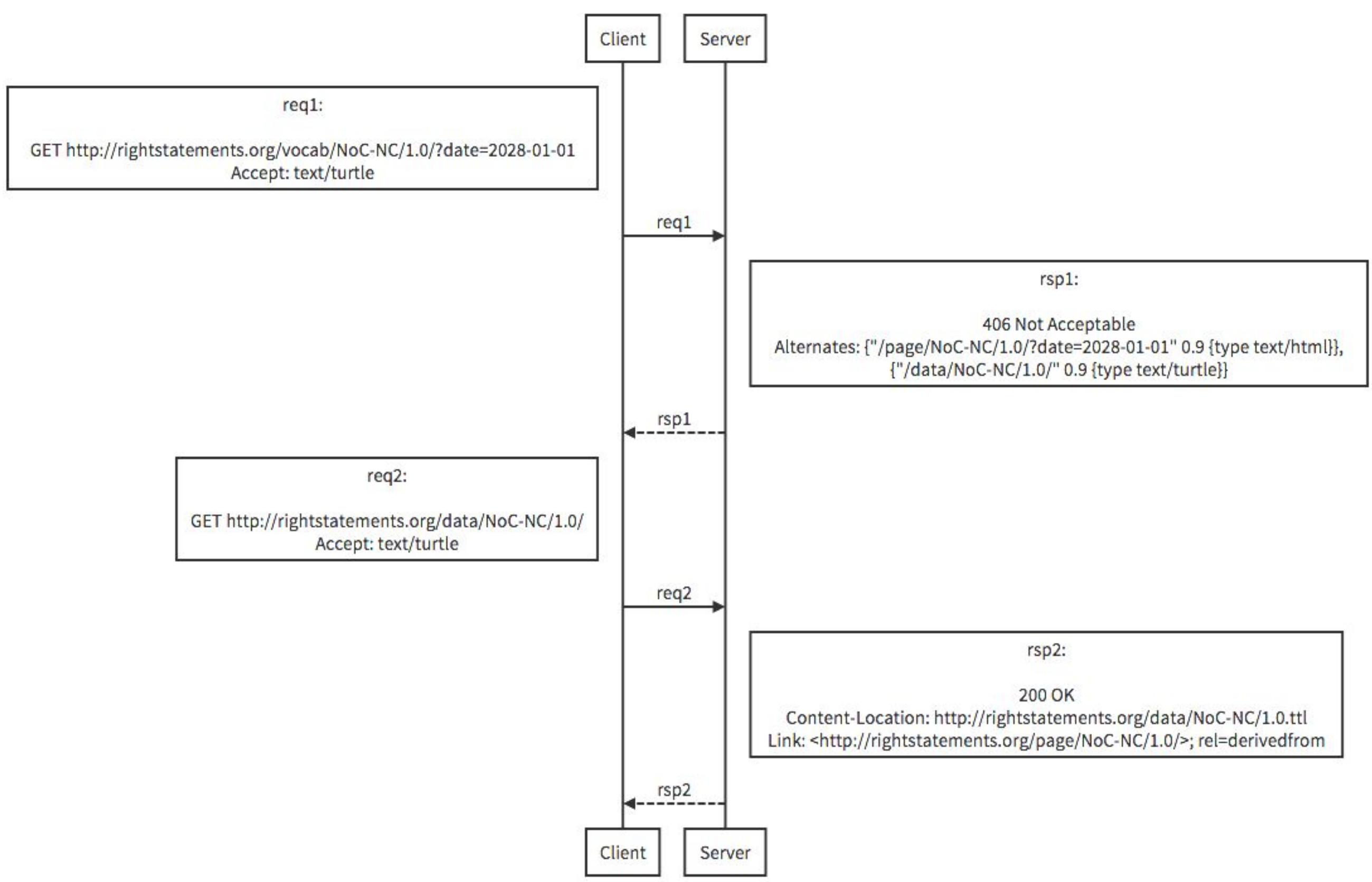

Dereference an NoC-NC statement with expiration by the Rights Statement URI, requesting HTML (invalid example, with “recovery” guidance):

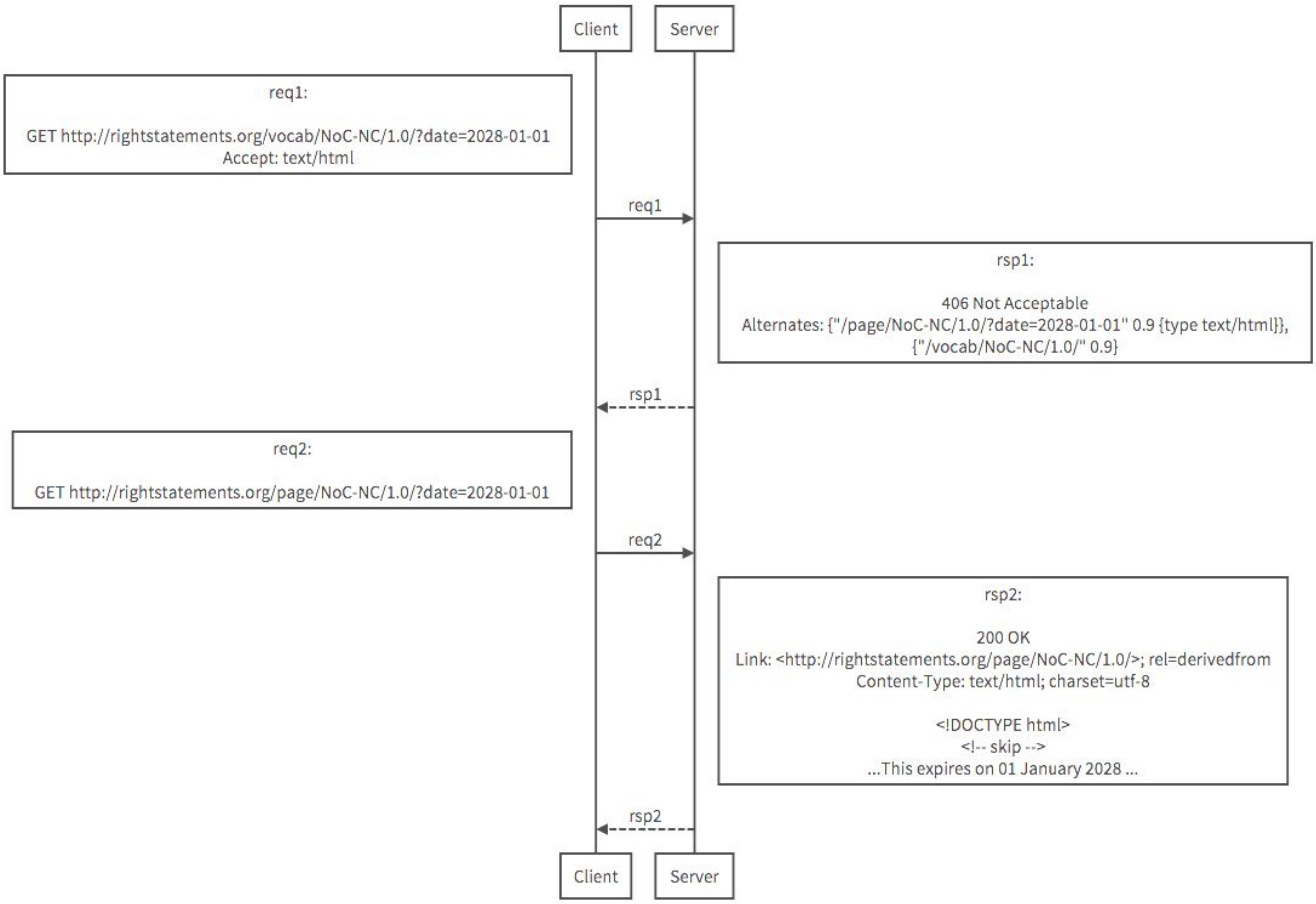

Dereference an InC statement with an invalid parameter by the representation URI, requesting HTML (invalid example, with “recovery” guidance):

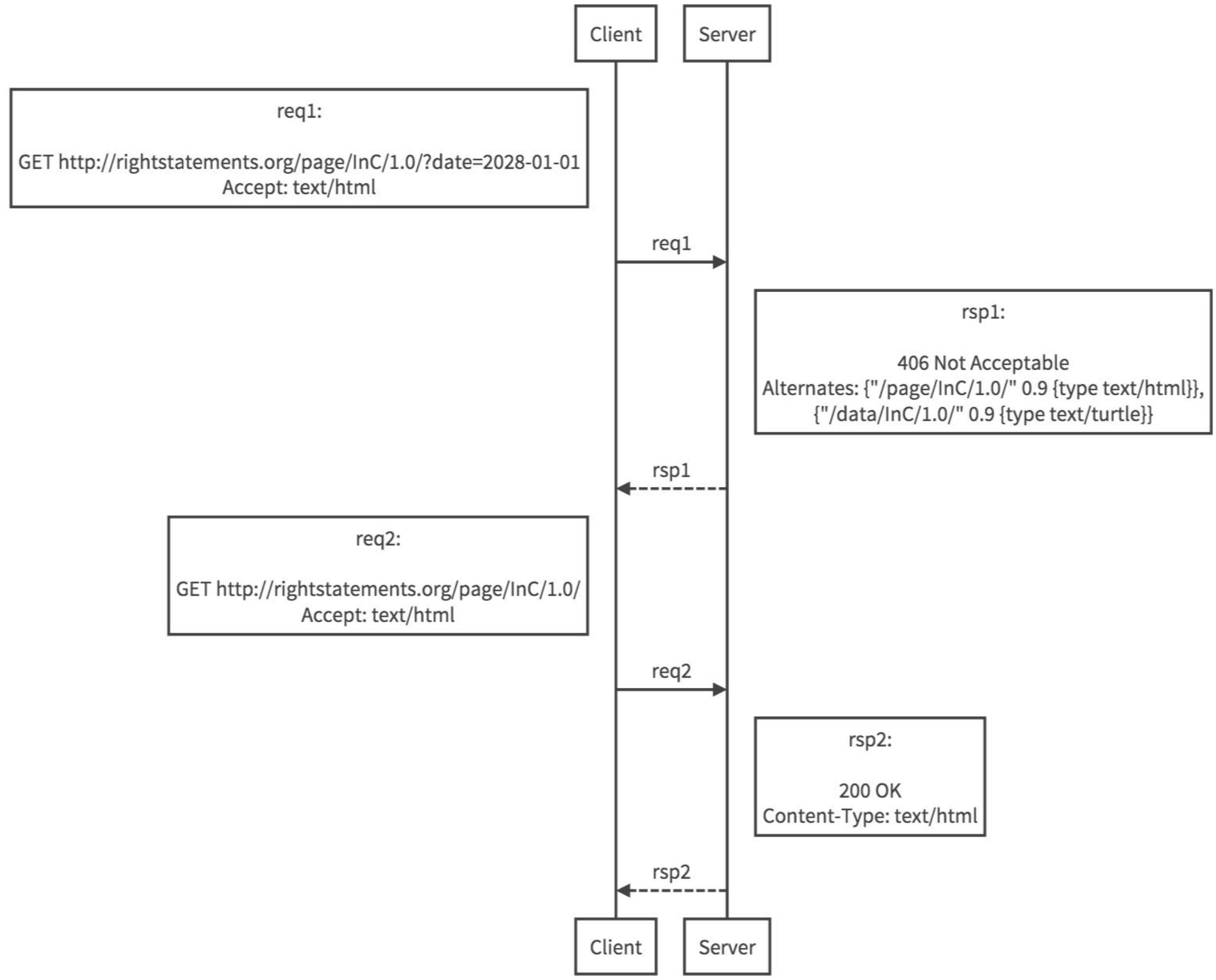

## Implementation guidelines

The proposed implementation is a custom web application, designed to serve the Rights Statements and their representations based upon the patterns listed above. The HTML representations for each translation is generated from the translations of literal values as expressed in the RDF vocabulary. The Technical Working Group investigated the use of existing software to publish RDF vocabularies, but found that they would not meet our user interface requirements without substantial modifications.

In terms of management of the vocabulary, we feel that a minimal infrastructure involving a version control system such as Git (e.g. hosting on GitHub) can satisfy the immediate needs. While we anticipate the vocabulary changing, the assumption is that changes will happen infrequently and through an orderly process. At this point, there are therefore no obvious technical requirements here. If more active, regular management is necessary, we propose the consideration of adopting a dedicated vocabulary management system.

# Acknowledgements

We are extremely grateful to the following individuals who contributed during the public feedback period for the *Recommendations for the Technical Infrastructure for Standardized Rights Statements*: Baxter Q. Andrews, Maarten Brinkerink, Nicholas Car, Leigh Dodds, Steven Folsom, Gloria Gonzalez, Kevin Hawkins, Jaffer, Lisette Kalshoven, Wibke Kolbmann, Sandra McIntyre, Aprille McKay, and Victor Rodriguez Doncel.